# ARACNE: An Algorithm for the Reconstruction of Gene Regulatory Networks in a Mammalian Cellular Context

Adam A. Margolin[1,2], Ilya Nemenman[2], Katia Basso[3], Chris Wiggins[2,4], Gustavo Stolovitzky[5], Riccardo Dalla Favera[3], Andrea Califano[1,2,*]

[1]Department of Biomedical Informatics, [2]Joint Centers for Systems Biology, [3]Institute for Cancer Genetics, [4]Department of Applied Physics and Applied Mathematics, Columbia University, New York, NY 10032

[5]IBM T.J. Watson Research Center, Yorktown Heights, N.Y. 10598

* Corresponding author:

1130 St. Nicholas Avenue Room 910, New York, NY 10032.

Email addresses:

AAM: adam@dbmi.columbia.edu, IN: ilya.nemenman@columbia.edu, KB: kb451@columbia.edu, CW: chw2@columbia.edu, GS: gustavo@us.ibm.com, RDF: rd10@columbia.edu, AC: califano@c2b2.columbia.edu

# Abstract

**Background**

Elucidating gene regulatory networks is crucial for understanding normal cell physiology and complex pathologic phenotypes. Existing computational methods for the genome-wide "reverse engineering" of such networks have been successful only for lower eukaryotes with simple genomes. Here we present *ARACNE*, a novel algorithm, using microarray expression profiles, specifically designed to scale up to the complexity of regulatory networks in mammalian cells, yet general enough to address a wider range of network deconvolution problems. This method uses an information theoretic approach to eliminate the majority of indirect interactions inferred by co-expression methods.

**Results**

We prove that ARACNE reconstructs the network exactly (asymptotically) if the effect of loops in the network topology is negligible, and we show that the algorithm works well in practice, even in the presence of numerous loops and complex topologies. We assess ARACNE's ability to reconstruct transcriptional regulatory networks using both a realistic synthetic dataset and a microarray dataset from human B cells. On synthetic datasets ARACNE achieves very low error rates and outperforms established methods, such as Relevance Networks and Bayesian Networks. Application to the deconvolution of genetic networks in human B cells demonstrates ARACNE's ability to infer validated transcriptional targets of the c-MYC proto-oncogene. We also study the effects of mis-estimation of mutual information on network reconstruction, and show that algorithms based on mutual information ranking are more resilient to estimation errors.

**Conclusions**

ARACNE shows promise in identifying direct transcriptional interactions in mammalian cellular networks, a problem that has challenged existing reverse engineering algorithms. This approach should enhance our ability to use microarray data to elucidate functional mechanisms that underlie cellular processes and to identify molecular targets of pharmacological compounds in mammalian cellular networks.

# Background

Cellular phenotypes are determined by the dynamical activity of large networks of co-regulated genes. Thus dissecting the mechanisms of phenotypic selection requires elucidating the functions of the individual genes in the context of the networks in which they operate. Because gene expression is regulated by proteins, which are themselves gene products, statistical associations between gene mRNA abundance levels, while not directly proportional to activated protein concentrations, should provide clues towards uncovering gene regulatory mechanisms. Consequently, the advent of high throughput microarray technologies to simultaneously measure mRNA abundance levels across an entire genome has spawned much research aimed at using these data to construct conceptual "gene network" models to concisely describe the regulatory influences that genes exert on each other.

Genome-wide clustering of gene expression profiles [1] provides an important first step towards this goal by grouping together genes that exhibit similar transcriptional responses to various cellular conditions, and are therefore likely to be involved in similar cellular processes. However, the organization of genes into co-regulated clusters provides a very coarse representation of the cellular network. In particular, it cannot separate statistical interactions that are irreducible (i.e., direct) from those arising from cascades of transcriptional interactions that correlate the expression of many non-interacting genes. More generally, as appreciated in statistical physics, long range order (i.e., high correlation among non-directly interacting variables) can easily result from short range interactions [2]. Thus correlations, or *any other* local dependency measure, cannot be used as the only tool for the reconstruction of interaction networks without additional assumptions.

Within the last few years a number of sophisticated approaches for the reverse engineering of cellular networks (also called deconvolution) from gene expression data have emerged (reviewed in [3]). Their goal is to produce a high-fidelity representation of the cellular network topology as a graph, where genes are represented as vertices and are connected by edges representing direct regulatory interactions. The criteria for defining an edge, as well as its biological interpretation, remain imprecise and vary between applications. For example, graphical modeling [4] defines edges as parent-child relationships between mRNA abundance levels that are most likely to explain the data, integrative methods [5] use independent experimental clues to define edges as those showing evidence of physical interactions, and other statistical / information theoretical methods [6] identify edges with the strongest statistical associations between mRNA abundance levels. All available approaches suffer to various degrees from problems such as overfitting, high computational complexity, reliance on non-realistic network models, or a critical dependency on supplementary data that is only available for simple organisms. These limitations have relegated the successful large-scale application of most methods to relatively simple organisms, such as the yeast *Saccharomyces cerevisiae*, and the genome-wide deconvolution of a mammalian network is yet to be reported.

Here we introduce *ARACNE* (Algorithm for the Reconstruction of Accurate Cellular Networks), a novel information-theoretic algorithm for the reverse engineering of transcriptional networks from microarray data that overcomes some of these limitations.

ARACNE defines an edge as an irreducible statistical dependency between gene expression profiles that cannot be explained as an artifact of other statistical dependencies in the network. We suggest that the presence of such irreducible statistical dependencies is likely to identify direct regulatory interactions mediated by a transcription factor binding to a target gene's promoter region, although other types of interactions may also be identified (see Discussion). In this study we focus on the former type of interaction for validation purposes. We demonstrate that ARACNE compares favorably with existing methods and achieves extremely low error rates in identifying transcriptional interactions in a synthetic dataset modeled using realistic Hill kinetics. In a biological context, we show that the algorithm infers bona-fide transcriptional targets in a mammalian gene network. We also study the effects of mis-estimation of mutual information (MI) on network reconstruction, and show that algorithms based on MI ranking are resilient to estimation errors. The algorithm is general enough to deal with a variety of other network reconstruction problems in biological, social, and engineering fields.

**Theoretical Background**

Several factors have impeded the reliable reconstruction of genome-wide mammalian networks. First, temporal gene expression data is difficult to obtain for higher eukaryotes, and cellular populations harvested from different individuals generally capture random steady states of the underlying biochemical dynamics. This precludes the use of methods that infer temporal associations and thus plausible causal relationships (reviewed in [7]). Only steady-state statistical dependences can be studied, which are not obviously linked to the underlying physical dependency model. Compounding this constraint, there is no universally accepted definition of statistical dependencies in the multivariate setting [8, 9]. In this work we adopt the definition of [9], which builds on ideas from the Markov networks literature [10]. Briefly, we write the joint probability distribution (JPD) of the stationary expressions of all genes, $P(\{g_i\})$, $i = 1,\ldots,N$, as:

$$P(\{g_i\}) = \frac{1}{Z}\exp\left[-\sum_i^N \phi_i(g_i) - \sum_{i,j}^N \phi_{ij}(g_i,g_j) - \sum_{i,j,k}^N \phi_{ijk}(g_i,g_j,g_k) - \ldots\right] \equiv e^{-H(\{g_i\})} \quad (1)$$

where $N$ is the number of genes, $Z$ is the normalization factor, also called the *partition function*, $\phi_{\ldots}$ are *potentials*, and $H(\{g_i\})$ is the *Hamiltonian* that defines the system's statistics. Within such a model, we assert that a set of variables interacts if and only if (*iff*) the single potential that depends exclusively on these variables is nonzero. ARACNE aims precisely at identifying which of these potentials are nonzero, and eliminating the others even though their corresponding marginal JPDs may not factorize. While this representation is not directly used by the algorithm, it helps precisely formalize our definition of interaction and the class of irreducible dependencies that it will help elucidate.

Note that Eq. (1) does not define the potentials uniquely, and additional constraints are needed to avoid the ambiguity (see Appendix B). A reasonable approach is to specify $\phi_{\ldots}$ using maximum entropy approximations [9, 11] to $P(g_1, \ldots, g_N)$ consistent with known marginals, so that constraining an *n*-way marginal defines its corresponding potential. We refer the reader to [9] for details.

**Approximations of the interaction structure**

Since typical microarray sample sizes are relatively small, inferring the exponential number of potential *n*-way interactions of Eq. (1) is infeasible and a set of simplifying assumptions must be made about the dependency structure. Eq. (1) provides a principled and controlled way to introduce such approximations. The simplest model is one where genes are assumed independent, i.e., $H(\{g_i\}) = \sum \phi_i(g_i)$, such that first-order potentials can be evaluated from the marginal probabilities, $P(g_i)$, which are estimated from experimental observations. As more data become available we should be able to reliably estimate higher order marginals and incorporate the corresponding potentials progressively, such that for $M \to \infty$ (where *M* is sample set size) the complete form of the JPD is restored. In fact, $M > 100$ is generally sufficient to estimate 2-way marginals in genomics problems, while $P(g_i, g_j, g_k)$ requires about an order of magnitude more samples. Thus the current version of ARACNE truncates Eq. (1) at the pairwise interactions level, $H(\{g_i\}) = \sum_i \phi_i(g_i) + \sum_{i,j} \phi_{ij}(g_i, g_j)$. Within this approximation, all genes for which $\phi_{ij} = 0$ are declared mutually non-interacting. This includes genes that are statistically independent (i.e., $P(g_i, g_j) \approx P(g_i)P(g_j)$), as well as genes that do not interact directly but are statistically dependent due to their interaction via other genes (i.e.. $P(g_i, g_j) \neq P(g_i)P(g_j)$, but $\phi_{ij} = 0$)[*].

Since the number of potential pairwise interactions is quadratic in *N*, identification of indirect statistical interactions is a formidable challenge for all network reconstruction algorithms that rely on statistical associations. However, under certain biologically realistic assumptions about the network topology, the ARACNE algorithm provides a framework to reconstruct two-way interaction networks reliably from a finite number of samples in a computationally feasible time.

# Algorithm

Within the assumption of a two-way network, all statistical dependencies can be inferred from pairwise marginals, and no higher order analysis is needed. While not implying that this is always the case for biological networks, it is important to understand whether this assumption may allow the inference of a subset of the true interactions with fewer false positives. Thus we identify candidate interactions by estimating pairwise gene expression profile mutual information, $I(g_i, g_j) \equiv I_{ij}$, an information-theoretic measure of relatedness that is zero *iff* $P(g_i, g_j) = P(g_i)P(g_j)$. We then filter MIs using an appropriate threshold, $I_0$, computed for a specific p-value, $p_0$, in the null-hypothesis of two independent genes. This step is basically equivalent to the Relevance Networks method [6] and suffers from the same significant limitations; namely, genes separated by one or more intermediaries (indirect relationships) may be highly co-regulated without implying an irreducible interaction, resulting in numerous false positives.

---

[*] $P(g_i, g_j) = P(g_i)P(g_j)$ is not a sufficient condition for $\phi_{ij} = 0$. We discuss this below.

Thus in its second step, ARACNE removes the vast majority of indirect candidate interactions ( $\phi_{ij} = 0$ ) using a well-known information theoretic property, the data processing inequality (DPI, discussed in detail later), that has not been previously applied to the reverse engineering of genetic networks.

**Mutual Information**

*Mutual information* for a pair of random variables, *x* and *y*, is defined as $I(x,y) = S(x) + S(y) - S(x,y)$, where $S(t)$ is the entropy of an arbitrary variable *t*. For a discrete variable, the *entropy* is $S(t) \equiv -\langle \log p(t_i) \rangle \equiv -\sum_i p(t_i) \log p(t_i)$ where $p(t_i) \equiv Prob(t = t_i)$ is the probability of each discrete state (value) of the variable (in this work, logarithms are natural). For continuous variables the entropy is infinite but the MI remains well defined and can be computed by replacing $S(x)$ with the *differential entropy*, which averages the log-probability density rather than the log-mass. Like the more familiar Pearson correlation, MI measures the degree of statistical dependency between two variables. However, while correlation coefficients are not invariant under reparameterizations and may be zero even for manifestly dependent variables, MI is reparameterization invariant and is nonzero *iff* any kind of statistical dependence exists.

*MI Estimation:* We estimate MI using a computationally efficient Gaussian Kernel estimator [12]. Given a set of two-dimensional measurements, $\vec{z}_i \equiv \{x_i, y_i\}, i = 1\ldots M$, the JPD is approximated as $f(\vec{z}) = 1/M \sum_i h^{-2} G\left(h^{-1} |\vec{z} - \vec{z}_i|\right)$, where $G(\ldots)$ is the bivariate standard normal density. With $f(x)$ and $f(y)$ being the marginals of $f(\vec{z})$, the MI is:

$$I(\{x_i\},\{y_i\}) = \frac{1}{M} \sum_i \log \frac{f(x_i, y_i)}{f(x_i) f(y_i)} \qquad (2)$$

Since MI is reparameterization invariant, we copula-transform (i.e., rank-order) [8] *x* and *y* for MI estimation; the range of these transformed variables is thus between 0 and 1, and their marginal probability distributions are manifestly uniform. This decreases the influence of arbitrary transformations involved in microarray data preprocessing and removes the need to consider position-dependent kernel widths, *h*, which might be preferable for non-uniformly distributed data.

For a spatially uniform *h*, the Gaussian kernel MI estimator is asymptotically unbiased for $M \to \infty$, as long as $h(M) \to 0$ and $[h(M)]^2 M \to \infty$. However, for finite $M$, the bias strongly depends on $h(M)$ and the correct choice is not universal. Fortunately, ARACNE's performance does not depend directly on the accuracy of the MI estimate, *I*, but rather on the accuracy of the estimation of MI ranks. For instance, determining if MI is statistically significant requires testing whether $I_{ij} \geq I_0$, where $I_0$ is the statistical significance threshold. Similarly, the DPI (see below) only requires ranking the MIs.

Producing reliable estimates of the MI ranks is an easier task. From the work on estimation of MI for discrete variables [13], we expect that, for well-sampled marginals and an undersampled joint, the bias is $b \approx b(\bar{I}, h)$ (where the bar denotes the true MI). Such biases almost cancel out for similar MI values; the ordering of MI estimates only weakly depends on the choice of $h$ and is stable even when MI itself is uncertain (**Figure 1**). Thus a single "ensemble best" value of $h$ can be used rather than optimizing the kernel width for each estimate (a computationally intensive operation). This result is general and should apply to any MI rank-based method. However, we emphasize that, since this result is largely empirical, the dependence of MI rank on the strength of smoothing should be reassessed for data sets with substantially different statistical properties before relying on this conclusion.

**Statistical Threshold for Mutual Information**

Since MI is always non-negative, its evaluation from random samples gives a positive value even for variables that are, in fact, mutually independent. Therefore, we eliminate all edges for which the null hypothesis of mutually independent genes cannot be ruled out. To this extent, we randomly shuffle the expression of genes across the various microarray profiles, similar to [6], and evaluate the MI for such manifestly independent genes and assign a p-value, $p$, to an MI threshold, $I_0$, by empirically estimating the fraction of the estimates below $I_0$. This is done for different sample sizes $M$ and for $10^5$ gene pairs so that reliable estimates of $I_0(p)$ are produced up to $p = 10^{-4}$. Extrapolation to smaller p-values is done using $p(I \geq I_0 | \bar{I} \quad 0) \propto e^{-\alpha M I_0}$, where the parameter $\alpha$ is fitted from the data. This formula is based on the intuition of the large deviation theory [14], which for discrete data and unbiased estimators suggests $p(I \geq I_0 | \bar{I} \quad 0) \propto e^{-M I_0}$. As MI in the continuous case can be estimated by finely discretizing the variables, a similar result should hold, and $\alpha$ accounts for possible biases of the estimator at fixed $h$. This produces an excellent agreement with numerical experiments (see supporting materials).

**Data Processing Inequality**

The DPI (**Figure 2**) [14] states that if genes $g_1$ and $g_3$ interact only through a third gene, $g_2$, (i.e., if the interaction network is $g_1 \leftrightarrow \ldots \quad g_2 \quad \ldots \quad g_3$ and no alternative path exists between $g_1$ and $g_3$), then

$$I(g_1, g_3) \leq \min\left[I(g_1, g_2); I(g_2, g_3)\right]. \tag{3}$$

Thus the least of the three MIs can come from indirect interactions only, and checking against the DPI may identify those gene pairs for which $\phi_{ij} = 0$ even though $P(g_i, g_j) \neq P(g_i) P(g_j)$. Correspondingly, ARACNE starts with a network graph where each $I_{ij} > I_0$ is represented by an edge $(ij)$. The algorithm then examines each gene triplet for which all three MIs are greater than $I_0$ and removes the edge with the smallest value. Each triplet is analyzed irrespectively of whether its edges have been marked for removal by prior DPI applications to different triplets. Thus the network reconstructed by the algorithm is independent of the order in which the triplets are examined.

Since this approach focuses only on the reconstruction of pairwise interaction networks, a pair of mutually independent genes, $I_{ij} < I_0$, will never be connected by an edge. Therefore, interactions represented by higher-order potentials for which the corresponding pairwise potentials are zero will not be recovered (see discussion). Additionally, even for a second order interaction network, one may imagine a situation where the effect of a direct interaction is exactly cancelled out by indirect interactions through other node(s), resulting in $\phi_{ij} \neq 0$ and $P(g_i, g_j) \approx P(g_i)P(g_j)$. This situation will not be identified by ARACNE. However, we believe that such precise cancellation is biologically unrealistic and the following theorems specify conditions under which ARACNE will reconstruct the network exactly.

Theorem 1[†]. If MIs can be estimated with no errors, then ARACNE reconstructs the underlying interaction network exactly, provided this network is a tree and has only pairwise interactions.

However, unlike standard tree reconstruction methods (e.g. Chow and Liu [15]), ARACNE is not limited to trees and can produce complicated structures containing many loops. In fact, because of the following two theorems, ARACNE can be viewed as a natural generalization of the Chow-Liu algorithm which overcomes the biologically-unrealistic tree assumption of the latter.

Theorem 2. The Chow-Liu (CL) maximum mutual information tree is a subnetwork of the network reconstructed by ARACNE.

Theorem 3. Let $\pi_{ik}$ be the set of nodes forming the shortest path in the network between nodes $i$ and $k$. Then, if MIs can be estimated without errors, ARACNE reconstructs an interaction network without false positives edges, provided: (a) the network consists only of pairwise interactions, (b) for each $j \in \pi_{ik}$, $I_{ij} \geq I_{ik}$. Further, ARACNE does not produce any false negatives, and the network reconstruction is exact *iff* (c) for each directly connected pair *(ij)* and for any other node $k$, we have $I_{ij} \geq \min(I_{jk}, I_{ik})$.

Tree networks satisfy all conditions of Theorem 3, while topologies containing loops may or may not. In particular, networks with three-gene loops definitely violate (c) [but may still satisfy (a) and (b)], and *every* such loop will be opened along the weakest edge. For a tree, there is a unique path that connects two nodes. Similarly, for networks that satisfy (a) and (b), the shortest path dominates inter-node information transfer. We call these networks *locally tree-like*. In other words, an interaction is retained by ARACNE if and only if there exist no alternate paths, via one or more intermediaries or branches on the network graph, which are a better explanation for the information exchange between two genes. Since biochemical dynamics is inherently stochastic, statistical interactions over more than a few separating edges are generically weak. Thus we believe that the local tree assumption is biologically realistic, and we expect ARACNE to produce low false positive rates in practice.

Finally, to minimize the impact of the variance of the MI estimator, a tolerance, $\tau$, may be introduced such that the DPI inequalities become of the form $I_{ij} \leq I_{ik}(1-\tau)$, and close values of MI are not pruned. For low values of $\tau$ (<15%) a reasonable tradeoff between true positives and false positives is achieved (see supporting materials). This threshold qualitatively matches the variance of the MI estimator and decreases with increasing sample size. Using such non-zero tolerance leads to persistence of some 3-gene loops.

---

[†] Proofs of all theorems can be found in the Appendix A.

**Algorithmic Complexity**

Because for a network of *N* genes there are at most *N* choose 3 gene triplets, ARACNE's complexity is $O(N^3 + N^2M^2)$, where *M* is the number of samples and *N* is the number of genes. The first term relates to the DPI analysis and the second to the mutual information estimation. This compares favorably with optimization methods that must explore an exponential search space (see Comparative Algorithms). In practice, the DPI is applied to a small subset of triplets for which all three edges survive the mutual information thresholding. Therefore, for large *M*, the computationally intensive part is generally associated with the second term (computing mutual information), which scales as $O(N^2M^2)$. As a result, ARACNE can efficiently analyze networks with tens of thousands of genes.

## Results

We study ARACNE's performance in reconstructing a class of synthetic networks proposed by [16] and a human B lymphocyte genetic network from gene expressions profile data. The latter has been reported in [17] and will only be recapitulated here. ARACNE's performance is compared against Relevance Networks (RNs) and Bayesian Networks (BNs). RNs are important to characterize the improvement associated with the introduction of the DPI, while BNs have emerged as some of the most widely used reverse engineering methods and provide an ideal comparative benchmark.

**Comparative Algorithms**

A *Bayesian Network* is a representation of a JPD as a directed acyclic graph (DAG) whose vertices correspond to random variables $\{X_1,\ldots,X_n\}$, and whose edges correspond to parent-child dependencies among variables; see [10] for an introduction and [18] for a more recent tutorial. We implemented the BN algorithm in this work in accordance with [19, 20]. In particular, we score graphs using the Bayesian scoring metric [21], for which we adopt a uniform prior over graphs and employ a Dirichlet prior over parameters to aid in the inference of undersampled conditional distributions of children given their parents. Such an approach inherently penalizes more complex graphs. Learning the most likely network requires exploring the entire graph space for the highest scoring model, which is an NP-complete problem [22]. Thus heuristic procedures are used to search for locally optimal graph structures. The comparative tests presented here use the greedy hill climbing algorithm with random restarts (simulated annealing and other structure search methods were tested and observed to produce similar results). These results were produced using the LibB software package [23], which is among the best implementations of the method.

*Relevance Networks* [6] compute mutual information for all gene pairs in a microarray dataset and infer that two genes are biologically related if their MI is above a certain threshold. This approach is equivalent to the first step in the ARACNE algorithm (i.e., without the DPI); however, we use a more accurate MI estimation procedure than the original implementation and have further developed the method of assigning statistical significance.

**Synthetic Networks**

*Networks Specification*: We benchmark the three algorithms using synthetic transcriptional networks proposed by Mendes *et al.* [16] as a platform for comparison of reverse engineering algorithms. These networks consist of 100 genes and 200 interactions organized either in an Erdös-Rényi (random network) [24] or a scale-free [25] topology (**Figure 3**). In the former, each vertex of a graph is equally likely to be connected to any other vertex; in the latter, the distribution of the number of connections, *k*, associated with each vertex follows a power law, $p(k) \sim k^{-\gamma}$ with $\gamma > 0$, and large interactions hubs are present. Many real biological networks appear to exhibit such structure [26].

The Mendes models use a multiplicative Hill kinetics to approximate transcriptional interactions:

$$\frac{dx_i}{dt} = a_i \prod_{j=1}^{N_I} \frac{IK_j^{n_j}}{IK_j^{n_j} + I_j^{n_j}} \prod_{l=1}^{N_A} \left(1 + \frac{A_l^{m_l}}{AK_l^{m_l} + A_l^{m_l}}\right) - b_i x_i , \qquad (4)$$

where $x_i$ is the concentration (expression) of the *i*-th gene, $N_I$ and $N_A$ are the number of upstream inhibitors and activators respectively, and their concentrations are $I_j$ and $A_l$. All other parameters are specified in [16].

We obtain synthetic expression values of each gene $x_i$ in each microarray $M_k$ by simulating its dynamics until the system relaxes to a steady state $\dot{x}_i \approx 0$. Prior to each simulation, the efficiency of synthesis and degradation reactions are varied by setting $a_i = \lambda_{k,i} \bar{a}_i$ and $b_i = \gamma_{k,i} \bar{b}_i$, where $\bar{a}_i$ and $\bar{b}_i$ are the original constant values of the parameters, and $\lambda_{k,i}, \gamma_{k,i}$ are random variables uniformly distributed in $[0.0, 2.0]$. Note that $\lambda_{k,i} \sim 0.0$ corresponds to a gene knock-out, while $\gamma_{k,i} \sim 2.0$ is a 2-fold increase in the synthesis rate. This parameter randomization models the sampling of a population of distinct cellular phenotypes at random time points (at or close to equilibrium), as is the case for the B cell experiments described later, where the efficiency of individual biochemical reactions may be different from assay to assay due to differences in temperature, nutrients, genetic mutations, etc. Although this model is a clear simplification of real biological networks, it forms a reasonably complex interaction network that captures some elements of transcriptional regulation, and an algorithm that does not perform well on this model is unlikely to perform well in a more complex case. Within this model, an interaction is unambiguously defined as a direct regulatory effect of one gene on another. Thus the performance of reverse engineering algorithms can be studied by comparing the inferred statistical interactions to the direct interactions in the model. We specifically note that, to our knowledge, this is the first attempt to benchmark network reverse engineering algorithms based on published objective criteria.

*Performance metrics*: Since genetic networks are sparse, potential false positives ($N_{FP}$), that is, identification of an irreducible statistical interaction between two genes not connected by a direct regulatory link, far exceed potential true positives ($N_{TP}$) [27]. Thus *specificity*, $N_{TN} / (N_{FP} + N_{TN})$, which is typically used in ROC analysis, is inappropriate as even small deviation from a value of 1 will result in large false positive numbers.

Therefore, we choose two closely related metrics, *precision* and *recall*. Recall, $N_{TP}/(N_{TP}+N_{FN})$, indicates the fraction of true interactions correctly inferred by the algorithm, while precision, $N_{TP}/(N_{TP}\ N_{FP})$, measures the fraction of true interactions among all inferred ones. Note that precision corresponds to the expected success rate in the experimental validation of predicted interactions. Performance will thus be assessed using Precision-Recall Curves (PRCs). PRCs for ARACNE and RNs are generated by adjusting the p-value or, equivalently, the MI threshold. ARACNE's PRC does not extend to 100% recall since the DPI eliminates some interactions even at $p_0 = 1$. To reach the 100% recall, the DPI tolerance, $\tau$, can be adjusted until ARACNE's PRC degenerates into that of RNs. For BNs, the adjustable parameter is the Dirichlet pseudocount, and, again, we observe that the maximum recall never reaches 100%.

*Performance Evaluation*: As shown in **Figure 4**, values of precision and recall for ARACNE are consistently better than those for the other tested methods. That is, for any reasonable precision (i.e. > 40%), ARACNE has a significantly higher recall than the other methods, and its precision reaches ~100% at significant recall levels. For large p-values, ARACNE begins to rapidly increase the number of false positives without a corresponding increase in true positives (the right tail of ARACNE's PRC). This is likely because as non-statistically significant MI values are accepted, random fluctuations may arbitrarily change the MI rank so that the DPI removes interactions at random. We note that the inflection of the PRC for ARACNE starts at $p_0 \sim 10^{-4}$, exactly where we would expect the algorithm to begin inferring large numbers of non-statistically significant interactions for a network of this size. This suggests that a sensible value for the MI threshold, producing a near optimal result, can be selected *a priori* using a Bonferroni-corrected p-value based on the number of potential network interactions.

ARACNE's high performance can be better understood by analyzing the distribution of MIs as a function of the length of the shortest path connecting each gene pair (degree of connectivity). ARACNE depends on MI being enriched for directly interacting genes and decreasing rapidly with this distance. **Figure 5** demonstrates these properties for our simulated datasets. There is no unique choice for the MI threshold that separates directly and indirectly interacting genes, and methods such as RNs that attempt to use a single threshold will either recover many indirect connections or miss a substantial number of direct ones. However, since mutual information decreases rapidly as signals travel over the network, the DPI effectively eliminates indirect interactions whose corresponding JPDs do not factorize. For all tested synthetic microarray sizes and both network topologies, ARACNE recovers far more true connections and far fewer false connections than the other methods (**Table 1**). Remarkably, in all cases, application of the DPI eliminates almost all indirect candidate interactions inferred by Relevance Networks at the expense of very few true interactions. We note that since ARACNE's performance degrades as the local topology deviates significantly from a tree, it performs slightly better on Erdös-Rényi than on scale-free topologies, for which small loops are more common. Another challenge in reconstructing the scale-free topology derives from the presence of large hubs with high in-degrees, which have small (and thus difficult to estimate) MI with their individual neighbors. However, ARACNE still performs extremely well even on scale-free topologies because signals in this network decorrelate rather quickly, so the statistical properties of a tree-like structure are locally preserved even in the presence of relatively tight loops (see Theorem 3). We note that ARACNE differs significantly from tree reconstruction methods, as the reconstructed topology for the scale-free network (using 1,000 samples) contains ~30 loops with sizes as small as four (see Appendix C for a description of our loop counting algorithm).

In summary, ARACNE appears to (a) achieve very high precision and substantial recall, even for few data points (125), (b) allow an optimal choice of the parameters h (Gaussian Kernel width) (**Figure 6**) and $I_0$ (statistical threshold), (c) to be quite stable with respect to the choice of parameters, and (d) to produce robust reconstruction of complex topologies containing many loops.

**Human B Cells**

Although large gene expression datasets such as those derived from systematic perturbations to simple organisms (e.g., [5]) are not easily obtained for mammalian cells, we suggest that an equivalent dynamic richness can be efficiently achieved by using a significant set of naturally occurring and experimentally generated phenotypic variations of a given cell type. To this end, we have assembled an expression profile dataset consisting of about 340 B lymphocytes derived from normal, tumor-related, and experimentally manipulated populations (for an extensive description see [28]).

This dataset was deconvoluted using ARACNE to generate a B cell specific regulatory network consisting of approximately 129,000 interactions. Since the c-MYC proto-oncogene emerges as one of the top 5% largest cellular hubs in the complete network and is extensively characterized in the literature as a transcription factor, we performed a first validation of the overall network quality by comparing its interactions inferred by our method with those previously identified by biochemical methods. The *in silico* generated network is highly enriched in known c-MYC targets; 29 out of 56 (51.8%) genes predicted to be first neighbors were either previously reported in the literature or biochemically validated in our labs, using chromatin immunoprecipitation, as c-MYC targets. This is statistically significant ($P = 2.9 \times 10^{-23}$ by $\chi^2$ test) with respect to the expected 11% of background c-MYC targets among randomly selected genes [29]. In addition, known c-MYC target genes were significantly more enriched among first neighbors than among second neighbors (51.8% vs. 19.4%), indicating that ARACNE is effective at separating direct regulatory interactions from indirect ones. Biological results related to the complete network structure are described in detail in [17].

# Discussion

ARACNE, which is motivated by statistical mechanics and based on an information theoretic approach, provides a provably exact network reconstruction under a controlled set of approximations. While we have shown that these approximations are reasonable even for complex mammalian gene networks, they may cause the algorithm to fail for some control structures. First, ARACNE will open all three-gene loops along the weakest interaction, and therefore introduce false negatives for triplets of interacting genes (although some may be preserved when a non-zero DPI threshold is used). Improvements to the algorithm are being investigated to address this condition. Second, by truncating Eq. (1) at the pairwise interactions, ARACNE will not infer statistical dependencies that are not expressed as pairwise interaction potentials (such as an XOR Boolean table for which MI between any gene pair is zero). By expanding Eq. (1) to include third and higher order potentials our formulation, in principle, can be extended to distinguish higher order interactions as well [30]. However, we note that in practice (i.e., biochemically) it is difficult to produce *only* higher order interactions without introducing some lower order dependencies [9], and truncation of the Hamiltonian is not likely to produce serious systematic errors in identifying interactions between gene pairs. In fact, the Mendes networks contain higher order interactions, but corresponding pairwise ones are effectively recovered instead. Another limitation of ARACNE is the inability to infer edge directionality, although we believe this to be a general limitation of all methods that do not use temporal data. We intend to investigate a two-tier approach in which first adirectional gene interactions are inferred, and then edge directionality is assessed via regression algorithms or specific biochemical perturbations.

Because mRNA abundance measurements only serve as a proxy for the interacting molecular species (i.e., activated protein concentrations), the type of physical interactions corresponding to the irreducible statistical dependencies identified by ARACNE are not always clear. For example, if the activity of a transcription factor is primarily mediated by an activating enzyme, rather than by changes in its mRNA abundance level, we expect ARACNE to identify dependencies between this enzyme and the target genes of the transcription factor. Moreover, a violation of the algorithm's hypotheses may occur for proteins involved in stable complex formation. Since it is energetically efficient for the cell to produce a stochiometrically balanced concentration of proteins involved in stable complexes (e.g., the ribosomal units), evolution has fine-tuned the transcriptional control of these proteins so that their concentrations are balanced. Thus, regardless of the concentration of the several transcription factors (TF) that may control their expression, the correlation between the final protein concentrations is generally higher than that between each protein and each individual TF. This violates the assumptions of Theorem 3 and produces irreducible statistical interactions between protein pairs involved in stable complex formation. Therefore, we expect some edges to correspond to protein-protein interactions, although we note that this situation would be correctly handled if higher order dependencies were analyzed.

Finally, we end with the following observation. Since ARACNE may fail for topologies with many tight loops, it is important to understand whether an analyzed topology is, in fact, locally tree like, and, therefore, the reconstruction can be trusted. We suggest two heuristics. First, loopy topologies continue to have more loops after reconstruction (results not shown). Thus an excessive number of loops in a deconvolved network should serve as a warning sign (Appendix C); more analysis is required to determine an acceptable range for this statistic. Second, as in the current analysis, predictions made by ARACNE (or, for that matter, any other computational algorithm) should be directly experimentally verified.

## Conclusions

The goal of ARACNE is not to recover *all* transcriptional interactions in a genetic network but rather to recover *some* transcriptional interactions with high confidence.

Within this scope, ARACNE overcomes several limitations that have impeded the application of previous methods to the genome-wide analysis of mammalian networks. It has a low computational complexity, does not require discretization of the expression levels, and does not rely on unrealistic network models or *a priori* assumptions. The algorithm can be applied to arbitrarily complex networks of transcriptional, or any other, interactions without reliance on heuristic search procedures. Thus we expect ARACNE to be well suited for mammalian gene regulatory networks, which are characterized by a complex topology, do not benefit from well-defined supplemental data (such as comprehensive protein interaction databases available for yeast), and are more difficult to manipulate experimentally, substantially hindering the acquisition of data to which time-series based methods can be applied. There are currently no other examples of a genome-wide mammalian network inferred from microarray expression profiles.

ARACNE's high precision in reconstructing a synthetic network designed to simulate transcriptional interactions, as well as the inference of bona-fide targets of c-MYC, a known transcription factor, in human B cells, suggest ARACNE's promise in identifying direct transcriptional interactions with low false-positive rates in mammalian networks, an obvious challenge for all reverse engineering algorithms. More research is needed to precisely characterize other types of interactions corresponding to irreducible statistical dependencies identified by ARACNE. We suggest that predictions made by ARACNE can be used in conjunction with other data modalities such as genome-wide location data, DNA sequence information, or targeted biochemical experiments to progress towards this level of detail. We plan to investigate this possibility using a model organism platform as well as extensions to the simulation model. However, studies based on targeted perturbations to model organisms have demonstrated the promise of using conceptual "gene-gene" networks to elucidate functional mechanisms underlying cellular processes [31] as well as to identify molecular targets of pharmacological compounds [32]. ARACNE may provide a framework to enable such applications in a mammalian context.

## Appendices

**Appendix A – Proofs of Theorems**

Theorem 1. If MIs can be estimated with no errors, then ARACNE reconstructs the underlying interaction network exactly, provided this network is a tree and has only pairwise interactions.

*Proof of Theorem 1.* First, notice that for every pair of nodes $i$ and $k$ not connected by a true direct interaction there is at least one other node $j$ that separates them on the network tree. Applying the DPI to the ($ijk$) triplet leads to removal of the ($ik$) edge. Thus only true edges survive. Similarly, every removed edge is not present in the true network. Consider some ($ijk$) triplet. One of these nodes, say $j$, may separate the other two. In this case the removed edge ($ik$) is clearly not in the true tree. Alternatively, there may be no separating node, and one may be able to move between any pair in the triplet without going through the third one. In this case none of the three edges is in the true graph, and any edge the DPI removes is fictitious. Thus all removed edges are indirect, while all remaining edges are factual. The network is reconstructed exactly.

Theorem 2. The Chow-Liu (CL) maximum mutual information tree is a subnetwork of the network reconstructed by ARACNE.

*Proof of Theorem 2.* We notice that, without a loss of generality, we can assume that the Chow-Liu tree and the ARACNE construction span all the nodes of the network. If this is not the case, that is, a few connected clusters exist (separated by edges with zero MI), then for the purpose of this theorem we can complete CL and ARACNE structures by the same edges with zero MI without formation of additional loops, till they become spanning. Now suppose that the theorem is false and there exists an edge *(ij)* that belongs to the (completed) CL tree, but does not belong to the ARACNE reconstruction. Since the CL construct is a tree, this edge separates it into two separate trees $T_i$ and $T_j$ that contain the *i*'th and the *j*'th nodes respectively. Since ARACNE has removed the *(ij)* link, there exists a node *k*, for which $\min(I_{ik}, I_{jk}) > I_{ij}$. Without a loss of generality, let *k* be in $T_i$. Then replacing the *(ij)* edge in the Chow-Liu tree by the *(jk)* edge will form no loops and will preserve the tree structure. This will increase the total MI of the CL reconstruction by $I_{jk} - I_{ij} > 0$. Thus the original tree is not the maximum MI tree. We arrive at a contradiction, which proves the theorem.

Theorem 3. Let $\pi_{ik}$ be the set of nodes forming the shortest path in the network between nodes *i* and *k*. Then, if MIs can be estimated without errors, ARACNE reconstructs an interaction network without false positives edges, provided: (a) the network consists only of pairwise interactions, (b) for each $j \in \pi_{ik}$, $I_{ij} \geq I_{ik}$. Further, ARACNE does not produce any false negatives, and the network reconstruction is exact *iff* (c) for each directly connected pair *(ij)* and for any other node *k*, we have $I_{ij} \geq \min(I_{jk}, I_{ik})$.

*Proof of Theorem 3.* To prove the absence of false positives, we notice that, for every candidate edge *(ik)* that is not actually in the network, there is at least one node *j*, such that $j \in \pi_{ik}$. Applying DPI to the *(ijk)* triplet will remove the *(ik)* edge. Further, we notice that if (c) is satisfied, then any application of DPI will not remove a true edge. However, if (c) does not hold, a true edge will be removed. This completes the proof.

**Appendix B - Relations to Graphical Models and Statistical Physics**

The definition of dependencies employed in the paper, which is based on the presence of a potential that couples interacting genes in the JPD,

$$P(\{g_i\}) = \frac{1}{Z} \exp\left[-\sum_i \phi_i(g_i) - \sum_{i,j} \phi_{ij}(g_i, g_j) - \sum_{i,j,k} \phi_{ijk}(g_i, g_j, g_k) - \cdots\right] \equiv e^{-H(\{g_i\})}, \quad (5)$$

is similar to that used in the theory of graphical models, specifically Markov Networks (MNs) [10]. However, even though there are some dissenting formulations (e.g., [33]), the usual implementation of MNs [10] is built using the notion of conditional (in)dependence. In this context it is impossible to distinguish, for example, a clique of three genes that are fully coupled by three pairwise interactions from the same genes coupled by a third order dependence, and also from a combination of both cases. Because of this, many authors use a convention that if a higher order potential $\phi_{...}$ is present in Equation 1, then all lower order potentials that depend only on a subset of the genes coupled by $\phi_{...}$ are incorporated into it. In contrast, the definition of [9], followed in this paper, aims at discriminating interaction orders. Thus, in our case, a three gene pairwise loop is distinct from a three-way interaction. In fact, extensions of ARACNE to deal with the latter have been developed [30], while the former still requires work.

As is understood in the graphical models literature, the formulation of Equation 1 resembles some statistical mechanics problems, specifically spin glasses on random networks [33, 34], particularly if the $g_i$ are binary (such discretization of expression levels is a common technique to deal with undersampling). In this case, the genes are the Ising spins, and truncations to the first, second, or the third order potentials are steps towards the mean field, Bethe, and Kikuchi variational approximations [33, 35-37]. An important distinction is that in statistical physics one searches for $\tilde{P}(\{g_i\})$, a variational approximation to the true JPD, $P(\{g_i\})$, that minimizes $D_{KL}(\tilde{P}\|P) \equiv \langle \log \tilde{P}/P \rangle_{\tilde{P}}$ within a given class of $\tilde{P}$, while the definition of [9] is equivalent to minimizing $D_{KL}(P\|\tilde{P})$. This is because statistical physics solves a direct problem – calculating various spin statistics given an interaction network. In particular, low order marginals $P_L$ are unknown and cannot be used in averaging. On the other hand, we are here solving the inverse problem – reconstructing the network given the known true marginal distributions.

ARACNE, which truncates Equation 1 at the second order potentials, is an analog of the Bethe approximation for the direct problem. Just like this approximation and the associated belief propagation algorithm [10, 38], ARACNE may fail for loopy topologies. It is, therefore, appealing that, for locally tree-like networks, the algorithm still works well, paralleling the corresponding discussion in statistical physics [38].

**Appendix C - Counting Loops in an Undirected Adjacency Matrix**

A pairwise interaction network can be represented by an adjacency matrix $A_{ij}$, where $A_{ij} = 1,0$ denotes either presence or absence of the corresponding interaction. To test the effect of violation of the "locally tree-like" assumption on the performance of the algorithm, we need to be able to count the number of cycles (loops) in a given network. This is complicated by the fact that the total number of cycles in a graph is not equal to the number of independent cycles; that is the number of edges that need to be removed to transform the graph into a tree. We need to count the number of independent cycles only. Additionally, of all possible complete sets of independent cycles we are interested in identifying the one with the smallest loops (since small loops have the highest potential to violate the locally tree-like assumption). We suggest the following algorithms to solve this task approximately.

1) We prune the nodes that have 0 or 1 neighbors in the adjacency matrix *A* (since such nodes cannot be part of any loops).

2) We transform the undirected network *A* into a directed one *B*. For this we identify every $A_{ij} \neq 0$ in the original network with a node in the new network (edges *ij* and *ji* are represented by separate nodes). If the original network had $A_{ij} = A_{jk} = 1,\ i \neq k$, then $B_{(ij),(jk)} = 1$, otherwise $B_{(ij),(kl)} = 0$.

3) We evaluate integer powers of the matrix *B*. If $Tr(B^n) > 0$, a loop (or loops) of size *n* are present. For the smallest *n* with loops, we identify one of them (at random), record nodes that form it, and remove one of these nodes in *B* (i.e., edges in *A*).

4) We repeat 1-3 till no more loops are found.

## Authors' contributions

AAM: Conducted research, designed study, participated in design of algorithm, wrote manuscript. IN: Designed theoretical framework, participated in design of algorithm, wrote manuscript. KB: Performed biochemical validation. CW: Participated in design of study. GS: Participated in design of algorithm and validation. RDF: Supervised and designed biochemical validation. AC: Designed algorithm, supervised research, wrote manuscript. All authors read and approved the final manuscript.

## Acknowledgements

This work was supported by the NCI (1R01CA109755-01A1) and the NIAID (1R01AI066116-01). AAM is supported by the NLM Medical Informatics Research Training Program (5 T15 LM007079-13).

# Figures

**Figure 1 – MI and MI rank estimation errors for varying Gaussian kernel widths**

The mean absolute percent error in estimating mutual information for bivariate normal densities is compared to the percent of errors in ranking the relative mutual information values for randomly sampled pairs for which the distribution with the lower true MI value is between 70% and 99% of the distribution with the higher value. MI estimation error (dashed blue line) is highly sensitive to the choice of Gaussian kernel width used by the estimator and grows rapidly for non-optimal parameter choices. However, due to similar bias for distributions with close MI values, the error in ranking pairs of MIs (solid red line) is much less sensitive to the choice of this parameter. These averages were produced using samples from 1,000 bivariate normal densities with a random uniformly distributed correlation coefficient $\rho \in [0.1, 0.9]$, such that $\bar{I} = -\frac{1}{2}\log(1-\rho^2)$. This results in a distribution of MI values that closely resembles that of the real microarray data.

**Figure 2 – Examples of the data processing inequality**

**(a)** $g_1$, $g_2$, $g_3$, and $g_4$ are connected in a linear chain relationship. Although all six gene pairs will likely have enriched mutual information, the DPI will infer the most likely path of information flow. For example, $g_1 \leftrightarrow g_3$ will be eliminated because $I(g_1,g_2) > I(g_1,g_3)$ and $I(g_2,g_3) > I(g_1,g_3)$. $g_2 \leftrightarrow g_4$ will be eliminated because $I(g_2,g_3) > I(g_2,g_4)$ and $I(g_3,g_4) > I(g_2,g_4)$. $g_1 \leftrightarrow g_4$ will be eliminated in two ways: first, because $I(g_1,g_2) > I(g_1,g_4)$ and $I(g_2,g_4) > I(g_1,g_4)$, and then because $I(g_1,g_3) > I(g_1,g_4)$ and $I(g_3,g_4) > I(g_1,g_4)$. **(b)** If the underlying interactions form a tree (and MI can be measured without errors), ARACNE will reconstruct the network exactly by removing all false candidate interactions (dashed blue lines) and retaining all true interactions (solid black lines).

**Figure 3 - Topology of the 100 gene regulatory networks proposed by Mendes**

Blue/red edges correspond to activation/inhibition. For the Erdös-Rényi topology **(a)** each gene is equally likely to be connected to every other gene, while the scale-free topology **(b)** is characterized by large interaction hubs with many connections.

**Figure 4 - Precision vs. Recall for 1,000 samples generated from the Mendes network**

**(a)** Erdös-Rényi network topology. **(b)** Scale-free topology. ARACNE's PRCs are consistently better than those of the other algorithms, and the precision reaches ~100% while maintaining high recall. Points on the PRCs for ARACNE and RNs corresponding to $p_0 = 10^{-4}$ (the value yieding $< 0.5$ expected false positives for 4,950 potential interactions) are indicated with arrows.

**Figure 5 - Distribution of mutual information for different lengths of the shortest path between genes for the scale-free topology**

Here we plot the log of the empirical probability that MI for a given separation between genes is above some value (in nats) marked on the horizontal axis. High MI values are significantly more probable for closer genes. Statistical significance threshold of $10^{-4}$ for the background MI distribution, corresponding to $I_0 = 0.0175$ nats, is marked on the graph. As shown, this threshold retains a large number of indirect candidate interactions, and there is no threshold that would be able to separate indirect and direct interactions; a threshold that eliminates most of the former (red arrows) also eliminates the majority of the latter. This severely degrades performance of RNs. (Inset) Expanded log-log view of the MI distribution for 934 gene pairs with 3 or more intermediaries and the background distribution computed by Monte Carlo. The curves are virtually indistinguishable, indicating that the background distribution can be used to obtain reliable estimates of statistical significance thresholds for filtering genes with higher degrees of connectivity.

connectivity. Similar results apply for the Erdös-Rényi topology (see supporting materials).

**Figure 6 – Synthetic network reconstruction errors for varying Gaussian kernel widths**

The total number of inferred errors $(N_{FP} + N_{FN})$ in reconstructing the Mendes networks is stable with respect to choice of estimator kernel width, validating the observation that rankings of MIs are more stable than the MI estimates with respect to changes in this parameter (**Figure 1**). The choice of kernel width for each number of samples that minimizes the mean absolute MI estimation error for bivariate Gaussian densities (indicated with diamonds) yields optimal or near optimal reconstruction of this network for all samples sizes. Results are calculated for a statistical significance threshold of $10^{-4}$ and a synthetic microarray size of 1,000 for the scale-free network topology.

## Tables

**Table 1**

Recovery for varying numbers of samples generated from the Mendes networks, which contain an average of ~194 true interactions after self-loops and bidirectional edges are eliminated. For all sample sizes ARACNE efficiently eliminates almost all false candidate interactions inferred by RNs, as indicated by the DPI sensitivity (calculated as the percent of false positives eliminated by the DPI), with minimal reduction in true positives, as indicated by the DPI precision (calculated as the percent of false positives removed out of the total number of edges removed by the DPI). Moreover, as the sample size decreases, the number of true connections inferred by ARACNE decays gracefully while the number of false positives remains very low, whereas the performance of Bayesian Networks degrades rapidly for smaller sample sizes as the conditional probability tables become very sparsely populated. Results are calculated using a p-value of $10^{-4}$ for ARACNE and Relevance Networks, yielding < 0.5 expected false positives for 4,950 potential interactions, and using a Dirichlet prior with equivalent sample size of one for Bayesian Networks [19]. Results are averaged over three network configurations for each topology.

# Figures

Figure 1.

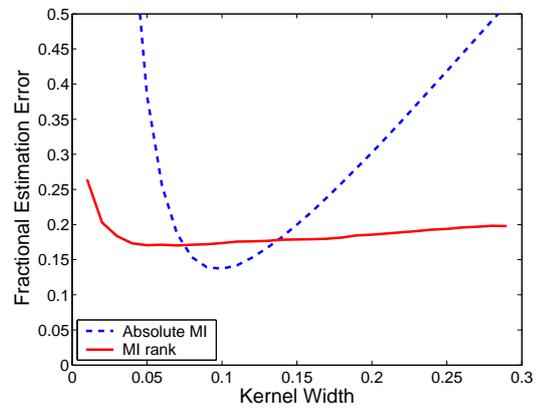

Figure 2.

(a) (b)

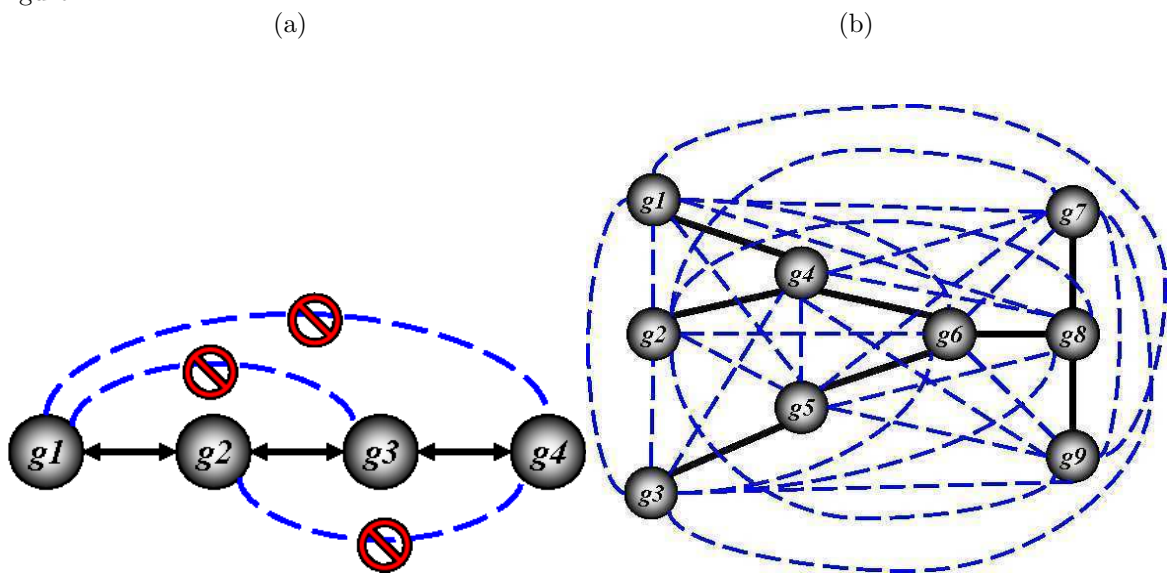

Figure 3.

(a) (b)

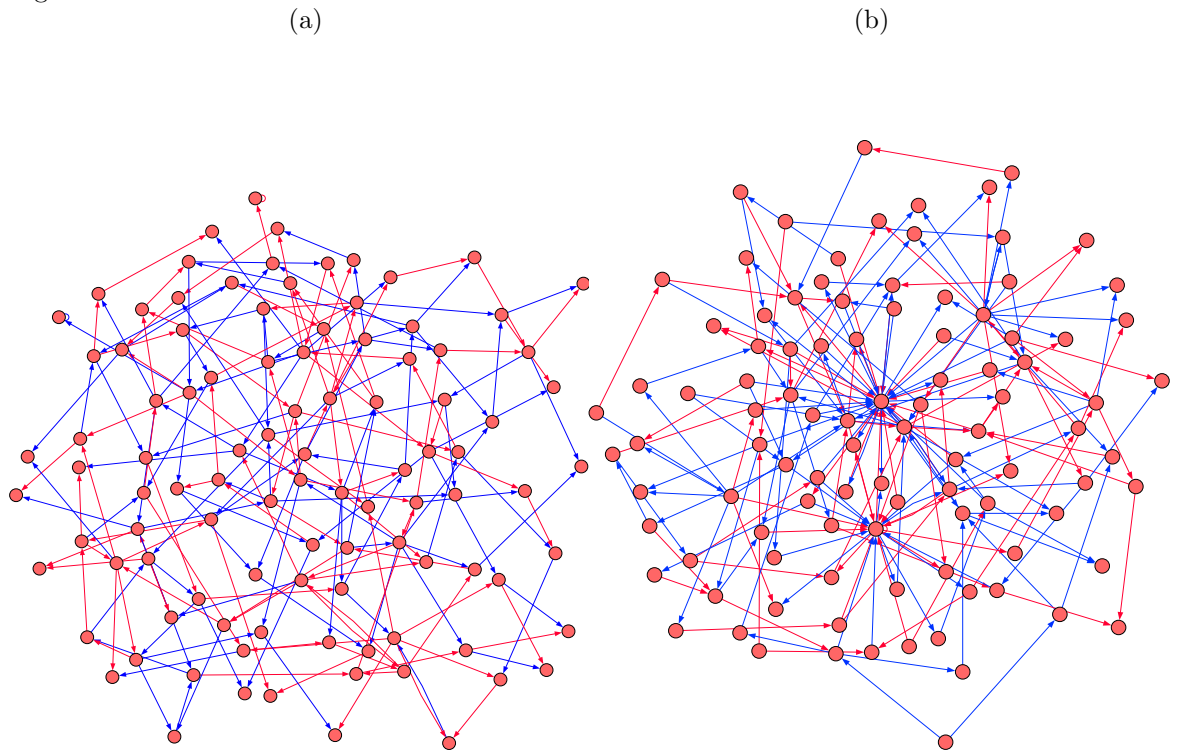

Figure 4.

(a) (b)

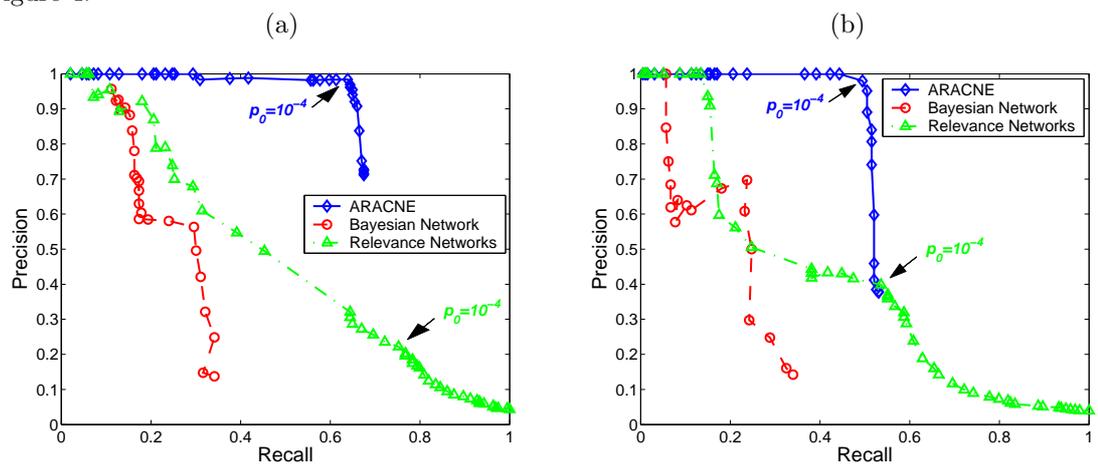

Figure 5.

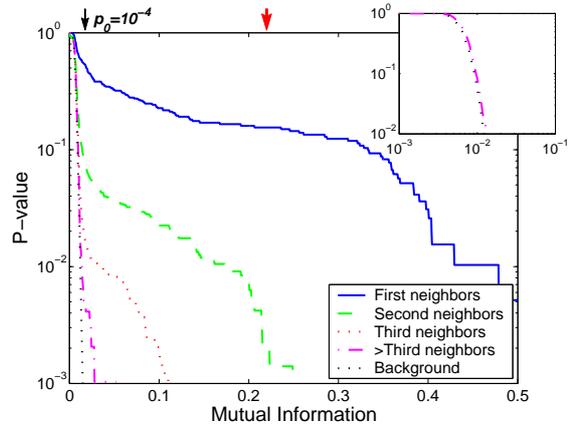

Figure 6.

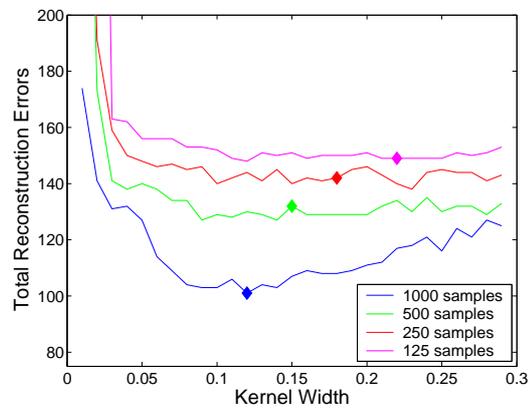

# Tables

Table 1.

***Erdös-Rényi Topology***

| Num samples | ARACNE | | Relevance Networks | | DPI Sensitivity | DPI Precision | Bayesian Networks | |
|---|---|---|---|---|---|---|---|---|
| | $N_{TP}$ | $N_{FP}$ | $N_{TP}$ | $N_{FP}$ | | | $N_{TP}$ | $N_{FP}$ |
| 1000 | 128.00 | 1.33 | 143.33 | 462.67 | 99.71% | 96.78% | 50.00 | 32.33 |
| 750 | 124.33 | 2.67 | 139.33 | 411.00 | 99.35% | 96.46% | 45.33 | 31.00 |
| 500 | 119.00 | 1.67 | 130.67 | 311.33 | 99.46% | 96.37% | 41.00 | 29.00 |
| 250 | 101.00 | 4.67 | 110.00 | 182.33 | 97.44% | 95.18% | 24.67 | 25.33 |
| 125 | 81.00 | 4.67 | 84.67 | 95.00 | 95.09% | 96.10% | 5.33 | 19.00 |

***Scale-Free Topology***

| Num samples | ARACNE | | Relevance Networks | | DPI Sensitivity | DPI Precision | Bayesian Networks | |
|---|---|---|---|---|---|---|---|---|
| | $N_{TP}$ | $N_{FP}$ | $N_{TP}$ | $N_{FP}$ | | | $N_{TP}$ | $N_{FP}$ |
| 1000 | 97.67 | 2.33 | 113.33 | 234.00 | 99.00% | 93.67% | 38.67 | 17.00 |
| 750 | 90.67 | 3.33 | 103.00 | 200.00 | 98.33% | 94.10% | 33.33 | 15.33 |
| 500 | 80.33 | 5.33 | 91.67 | 154.67 | 96.55% | 92.95% | 27.00 | 13.33 |
| 250 | 63.33 | 7.67 | 70.00 | 80.00 | 90.42% | 91.56% | 9.00 | 9.67 |
| 125 | 46.33 | 3.67 | 48.00 | 49.67 | 92.62% | 96.50% | 4.00 | 6.00 |

# Supporting Figures Legends

**Figure 7**
P-values are assigned to MI thresholds using a Monte Carlo simulation for different kernel widths, sample sizes ($M$) and for $10^5$ gene pairs so that reliable estimates are produced up to $p = 10^{-4}$ (solid lines). Extrapolation to smaller p-values is done using $p(I \geq I_0 \mid \bar{I} = 0) \propto e^{-\alpha M I_0}$ (dotted lines).

**Figure 8**
The number of inferred errors, $N_{FP} + N_{FN}$, are plotted as a function of the DPI tolerance, $\tau$, for **(a)** the Erdös-Rényi and **(b)** the scale-free topologies. Raising $\tau$ to a value of 0.2 results in a modest increase in false positives, while larger values of $\tau$ produce a much sharper increase. Therefore, a moderate choice for the tolerance can help elucidate additional interactions without introducing an excessive number of false positives. Results are calculated for a statistical significance threshold of $10^{-4}$ and a synthetic microarray size of 1,000.

**Figure 9**
Distribution of mutual information for different lengths of the shortest path between genes for the Erdös-Rényi topology. Red and black arrows are explained in the legend of Figure 5. Since there are no large in-degree hubs, decorrelation is slower than for the scale-free network, and MI statistics even for fifth neighbors is still distinguishable from the background.

# Supporting Figures

Supporting Figure 7.

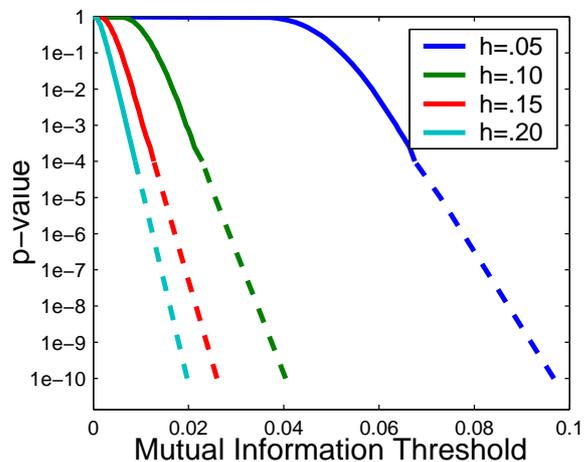

Supporting Figure 8.

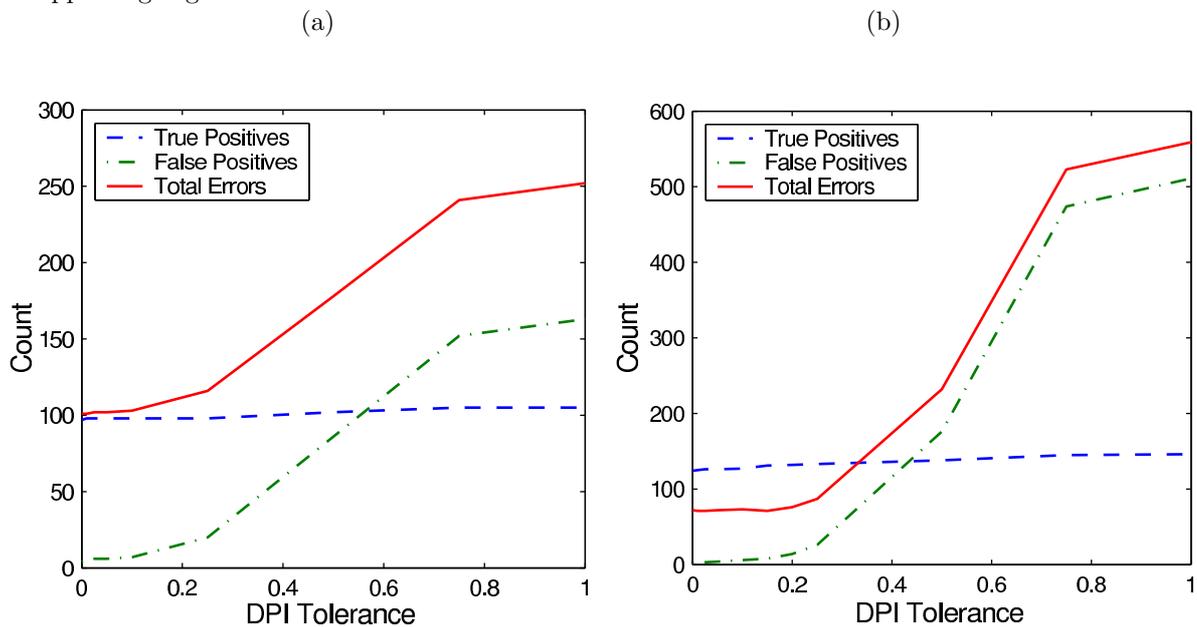

Supporting Figure 9.

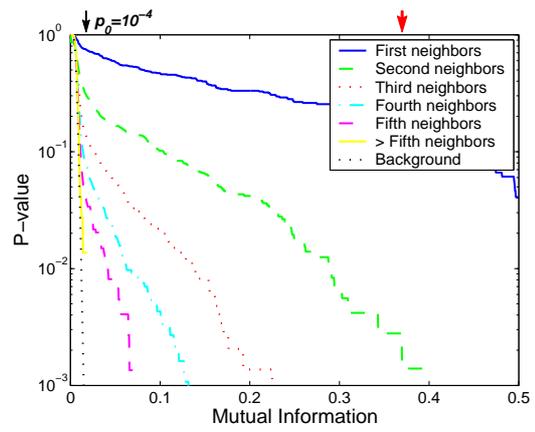